\documentclass[aps,pra,letterpaper,twocolumn,superscriptaddress,floatfix]{revtex4-2}

\usepackage{eqnarray}
\usepackage{amsmath}    
\usepackage{graphicx}   
\usepackage{verbatim}   
\usepackage{color}      
\usepackage{subfigure}  
\usepackage{hyperref}   
\usepackage{physics}
\usepackage{dsfont}
\usepackage{amsmath}
\usepackage{amssymb}
\usepackage{mathrsfs}
\usepackage{nccmath}
\usepackage{microtype}
\usepackage{algorithm2e}

\begin{document}


\title{Finding the Dynamics of an Integrable Quantum Many-Body System via Machine Learning}

\author{Victor Wei}
\email{yanfei.wei@mail.mcgill.ca}
\author{Alev Orfi}
\email{alev.orfi@uwaterloo.ca}
\affiliation{Department of Physics, McGill University, Montreal, QC, Canada}
\affiliation{Institute for Quantum Computing, University of Waterloo, Waterloo, ON, Canada}
\affiliation{Department of Physics and Astronomy, University of Waterloo, ON, Canada}
\author{Felix Fehse} 
\email{felix.fehse@mail.mcgill.ca}
\author{W.~A.~Coish}
\email{william.coish@mcgill.ca}
\affiliation{Department of Physics, McGill University, Montreal, QC, Canada}

\begin{abstract}
We study the dynamics of the Gaudin magnet (``central-spin  model'') using machine-learning methods. This model is of practical importance, e.g., for studying non-Markovian decoherence dynamics of a central spin interacting with a large bath of environmental spins and for studies of nonequilibrium superconductivity. The Gaudin magnet is also integrable, admitting many conserved quantities: For $N$ spins, the model Hamiltonian can be written as the sum of $N$ independent commuting operators. Despite this high degree of symmetry, a general closed-form analytic solution for the dynamics of this many-body problem remains elusive. Machine-learning methods may be well suited to exploiting the high degree of symmetry in integrable problems, even when an explicit analytic solution is not obvious. Motivated in part by this intuition, we use a neural-network representation (restricted Boltzmann machine) for each variational eigenstate of the model Hamiltonian. We then obtain accurate representations of the ground state and of the low-lying excited states of the Gaudin-magnet Hamiltonian through a variational Monte Carlo calculation. From the low-lying eigenstates, we find the non-perturbative dynamic transverse spin susceptibility, describing the linear response of a central spin to a time-varying transverse magnetic field in the presence of a spin bath. Having an efficient description of this susceptibility opens the door to improved characterization and quantum control procedures for qubits interacting with an environment of quantum two-level systems. These systems include electron-spin and hole-spin qubits interacting with environmental nuclear spins via hyperfine interactions or qubits with charge or flux degrees of freedom interacting with coherent charge or paramagnetic impurities.
\end{abstract}
\date{\today}
\maketitle


\section{Introduction}
Predicting the dynamics of quantum many-body systems is crucial for understanding many important physical phenomena. For example, the dynamics of the Fermi-Hubbard model can advance our understanding of superconductivity and quantum magnetism in correlated materials \cite{lee2006doping, imada1998metal}. However, brute-force numerical approaches such as exact diagonalization on a classical computer have an exponential cost in time and/or memory and can therefore only be used to simulate small quantum systems. To tackle the problem of quantum many-body simulation, a large-scale fault-tolerant quantum computer could be used to efficiently run quantum simulations with predictable bounded errors \cite{peruzzo2014variational,lloyd1996universal}. The hardware challenge behind building a useful quantum computer is, however, significant. General-purpose quantum simulation on a quantum computer may not be feasible until far in the future. 

Despite the limitations of classical computers in simulating quantum systems, there are nevertheless many classical algorithms that are effective in special cases. A notable example is the variational Monte Carlo (VMC) method, which finds an approximate ground state upon minimizing the estimated energy with respect to a variational ansatz \cite{foulkes2001quantum}. Recently, Carleo et al.~have used a neural-network ansatz in VMC to calculate ground states of many-body Hamiltonians with a better accuracy than other existing methods \cite{carleo2017solving}. A number of follow-up works on neural-network quantum states also strongly suggest that neural networks are a promising model for representing a large subset of quantum states and for approximating the ground states of many important many-body Hamiltonians \cite{bennewitz2022neural,carrasquilla2019reconstructing,chen2018equivalence,kiyohara2020learning,choo2018symmetries, melko2019restricted}. Using this type of ansatz, it is also possible to leverage well-studied optimization strategies from the deep-learning community.
\begin{figure}
    \centering
    \includegraphics[width=0.95\columnwidth]{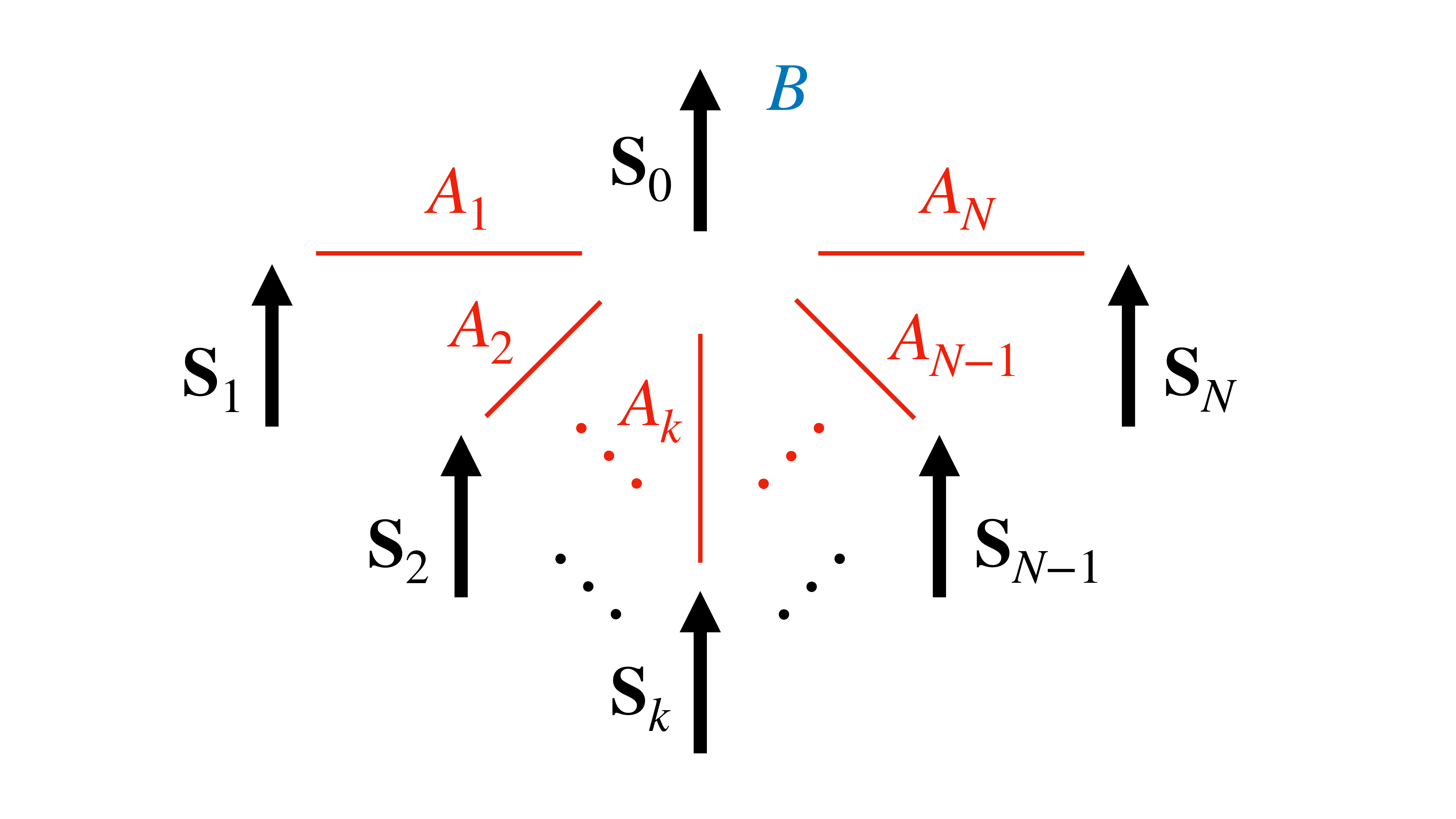}
    \caption{Diagrammatic representation of the Gaudin magnet, where $\boldsymbol{S}_k$ are the individual spin-1/2 operators, $A_k$ is the coupling between the $k^\mathrm{th}$ spin and the central spin, and $B$ is the external field.}
    \label{fig:cs}
\end{figure}

A number of recent works have extended neural-network methods from the realm of static properties to many-body quantum dynamics \cite{flurin2020using, carleo2017unitary, herrera2021convolutional, medvidovic2021classical, schmitt2022quantum}. These have demonstrated success in some test cases (often where an analytic solution is already available), but they are unlikely to be successful for any arbitrary problem \cite{childs2014bose, wei2010interacting, huang20202d}. Finding a subclass of problems that have known practical applications, that are nontrivial (where no closed-form analytic solution is available), and that have a good chance of admitting an efficient neural-network solution is an important open question.  Here, we apply machine-learning methods to determine the low-energy spectrum and non-perturbative low-frequency dynamic response for one such model: the Gaudin magnet (\textbf{Figure \ref{fig:cs}}). This model applies directly to the many-body decoherence dynamics of a ``central-spin'' problem (describing, e.g., an electron spin qubit interacting with a bath of nuclear spins) \cite{khaetskii2002electron, schliemann2003electron, coish2004hyperfine, fischer2008spin, bortz2007exact, bortz2010dynamics, bortz2010spectrum, barnes2012nonperturbative,stanek2013dynamics,stanek2014quantum,lindoy2018simple}. A linear combination of commuting Gaudin magnets can also result in the Bardeen-Cooper-Schrieffer (BCS) pairing Hamiltonian describing non-equilibrium superconductivity, with fermionic creation/annihilation operators represented in terms of Anderson pseudospins \cite{cambiaggio1997integrability, sierra2002integrability, asorey2002chern,ruh2023digital}. A complete solution for the dynamics of the Gaudin magnet thus implies a solution for BCS pairing dynamics (through the composition of a product of commuting unitary time-evolution operators, one for each Gaudin-magnet problem). The Gaudin magnet is an integrable system, with a large number of conserved quantities. This suggests that drastic simplifications are possible. Although these simplifications are difficult to realize through direct analytic study, a well-designed machine-learning procedure may be able to exploit the high degree of symmetry in this problem. 

For integrable $N$-particle systems, the $O(e^N)$ set of linear equations that would normally describe the Schr\"odinger equation in a complete basis can be recast into a set of $O(N)$ nonlinear Bethe ansatz equations \cite{takhtajan1985introduction}. The Gaudin magnet is one such system, and its integrability has been exploited in a number of works to obtain eigenstates and observable dynamics \cite{bortz2007exact,faribault2013integrability,faribault2013spin,he2022quantum}. Despite the reduction to a much smaller system of Bethe ansatz equations, solving these nonlinear equations is still nontrivial in general. For example, a hybrid method based on a combination of the algebraic Bethe ansatz and direct Monte Carlo sampling proposed by Faribault and Schuricht is still limited to modest system sizes ($N\lesssim 50$) \cite{faribault2013integrability}. 

Knowing that the eigenstates of the Gaudin magnet can be derived from $O(N)$ nonlinear equations, we view a neural-network quantum state ansatz as a promising candidate for obtaining the low-lying eigenstates and dynamics of this system. Neural networks such as the restricted Boltzmann machine (RBM)  \cite{zhang2018overview} can be used to describe quantum states using a number of network parameters that grows only polynomially in the system size $N$, but they can nevertheless be used to describe complex quantum states through the nonlinearity induced by tracing over hidden layers. We speculate that neural-network methods may be able to calculate the low-lying eigenstates by learning the symmetries of the integrable Gaudin magnet and other integrable models. 

In this work, we calculate the low-lying eigenstates of the Gaudin magnet (central-spin model). To do this, we use a penalty-based variational algorithm with an RBM neural-network ansatz. We also compute the non-perturbative transverse spin susceptibility for the central spin, giving  the linear response of the central spin to a time-varying magnetic field, accounting for highly accurate representations of the many-body eigenstates of the central spin interacting with environmental spins. We find these approximate eigenstates (and the associated transverse spin susceptibility) numerically in a regime where conventional perturbation theory fails.

Due to the widespread practical importance of the Gaudin magnet, there have been many other proposed strategies to determine the non-perturbative dynamics of this model. The most natural approach is to restrict to uniform coupling coefficients (the box model), in which case the problem can be diagonalized exactly analytically \cite{zhang2006hyperfine, coish2007quantum}. The box model can accurately describe dynamics for the original problem (with a broad distribution of coupling coefficients) at short times (where semiclassical methods also apply). The box model is, however, too restrictive to be used to understand long-time quantum dynamics in most practical contexts, except where uniform coupling coefficients can be engineered directly \cite{chesi2015theory}. Several authors have considered an extension to a so-called wedding-cake model, composed of layered rings of environmental spins that behave as single larger spins having equal coupling constants \cite{ramon2007dynamical,fang2021superradiantlike,vezvaee2021driven}. The wedding-cake model allows for efficient exact numerical calculation while sacrificing the original distribution of coupling coefficients. Recent work has analyzed the case where the distribution of piecewise-constant coupling coefficients is chosen to reproduce moments of the original coupling-constant distribution \cite{lindoy2018simple}. This method accurately reproduces correlation functions at infinite temperature and at full polarization, but it is not clear how this wedding-cake method will perform for low-temperature and non-equilibrium initial states, as considered here. Finally, numerical methods based on the time-dependent density-matrix renormalization group \cite{stanek2013dynamics,stanek2014quantum} and analytic methods based on a partial resummation \cite{barnes2012nonperturbative} are both limited to short time scales (set by the largest coupling coefficient), while we will be interested in dynamics at and beyond this time scale in the present work.  

The rest of this paper is organized as follows: In Section \ref{sec:Model}, we introduce the Gaudin-magnet model Hamiltonian and describe a particular physically relevant set of coupling constants. In Section \ref{sec:VariationalAlgorithms}, we describe the variational ansatz and variational algorithms used to find accurate representations of the ground state and several low-lying excited states. Section \ref{sec:Dynamics} describes the dynamic transverse spin susceptibility for this problem and presents sample calculations for the non-perturbative spectral function and linear response to a low-frequency transverse field. Section \ref{sec:Conclusions} summarizes the potential advantages and shortcomings of this approach, along with conclusions and possible future directions. A technical derivation of the gradient expression used for gradient-based optimization is given in the Appendix.
\section{Model}\label{sec:Model}
The central-spin model (Gaudin magnet) consists of one central spin coupled to $N$ environment spins (Figure \ref{fig:cs}) \cite{gaudin1976diagonalisation,schliemann2003electron,yuzbashyan2005solution}. The central spin is coupled to the $k^\mathrm{th}$ environment spin through a Heisenberg interaction with coupling coefficient $A_k$. The central spin alone is additionally subject to a constant external field $B$. The Hamiltonian for the Gaudin magnet is thus
\begin{equation}
H=BS_0^z+\sum_{k=1}^{N}A_k\mathbf{S}_0\cdot\mathbf{S}_k,
\label{Hamiltonian}
\end{equation}
where we have set $\hbar=1$, $S_j^{\alpha}=\frac{1}{2}\sigma_j^\alpha$ ($\alpha=x,y,z$) is the spin-1/2 operator for spin $j$ ($j=0$ for the central spin and $j=k=1,2,\ldots,N$ for the environment spins). The operator $\sigma_j^\alpha$ ($\alpha=x,y,z$) is the Pauli operator for spin $j$. 

\begin{figure}
    \centering
    \includegraphics[width=0.95\columnwidth]{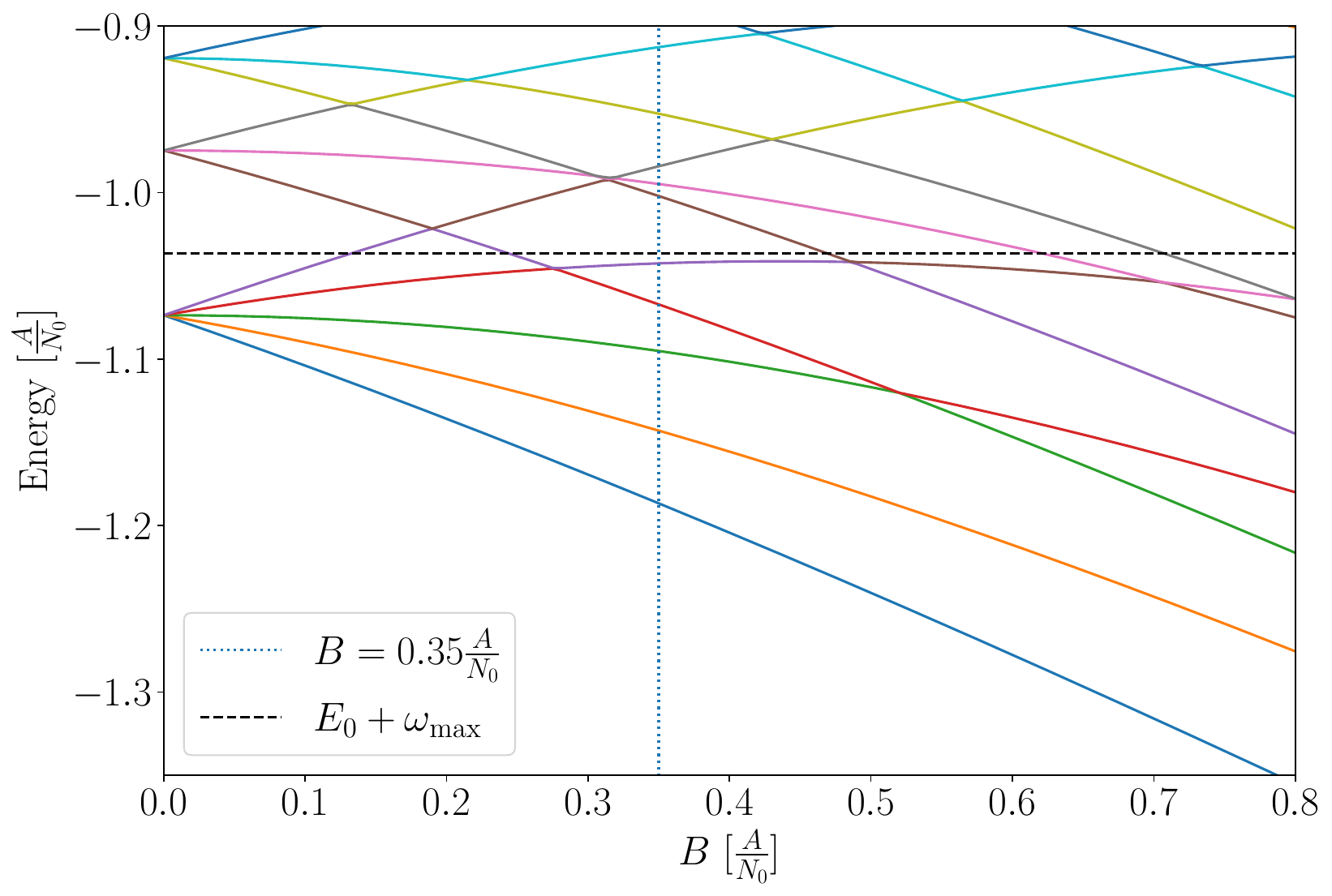}
    \caption{Energy spectrum of the Gaudin magnet (central spin model) with $N=5$ environment spins and an exponential distribution of coupling coefficients [Equation \eqref{Ak}] with range $N_0=3$. The vertical dashed line indicates the value $B = 0.35 A/N_0$ used in dynamics calculations. The horizontal dashed line shows $E_0 + \omega_{\mathrm{max}}$, with $E_0$ the ground-state energy for $B=0.35 A/N_0$ and with $\omega_{\mathrm{max}} = 0.15 A/N_0$. The parameter $\omega_\mathrm{max}$ sets the maximum frequency cutoff in calculating the low-frequency dynamical response.}
    \label{fig:level}
\end{figure}

In numerical evaluations, we choose the coupling coefficients $A_k$ to have an exponential distribution,
\begin{equation}
A_k = \frac{A}{N_0}e^\frac{-(k-1)}{N_0}.
\label{Ak}
\end{equation}
The total number of environment spins in the model is $N$ while $N_0$ controls the scale where the coupling coefficients decay exponentially. We will typically be interested in systems where $N>N_0$; in numerical evaluations below, we choose $N_0=(N+1)/2$. An exponentially decaying coupling strength corresponds, e.g., to the distribution of hyperfine couplings expected for an electron spin (central spin $\mathbf{S}_0$) interacting with nuclear spins (environment spins $\mathbf{S}_k$) in a two-dimensional quantum dot with parabolic confinement \cite{coish2004hyperfine}. In this example, $N_0$ is the number of nuclear spins within a quantum-dot Bohr radius, while $N\to\infty$ is the number of nuclear spins in the entire crystal.  The low-lying spectrum of this model is shown for $N=5$ [$N_0=(N+1)/2=3$] in \textbf{Figure \ref{fig:level}} for a range of $B$. In dynamics calculations below, we set $B=0.35 A/N_0$ (blue dashed line in Figure \ref{fig:level}), chosen to avoid degeneracies up to the fourth excited state. For $B\gg A$, we can directly apply perturbation theory to give the eigenstates of $H$ as the simultaneous eigenstates of all $S^z_j$ and in this case the spectrum is trivial. 
However, for $B\lesssim A$, this simple perturbation theory does not generally apply and more advanced methods are required to understand the detailed spin dynamics arising from this model.

\section{Variational Algorithms}\label{sec:VariationalAlgorithms}
In this section, we introduce the RBM variational ansatz used to approximate the eigenstates of the Hamiltonian $H$. We then describe the variational algorithms used to optimize the RBM network parameters and we explain strategies used to select accurate representations of the states.

\subsection{Variational Ansatz}
\begin{figure}
    \centering
    \includegraphics[width=0.95\columnwidth]{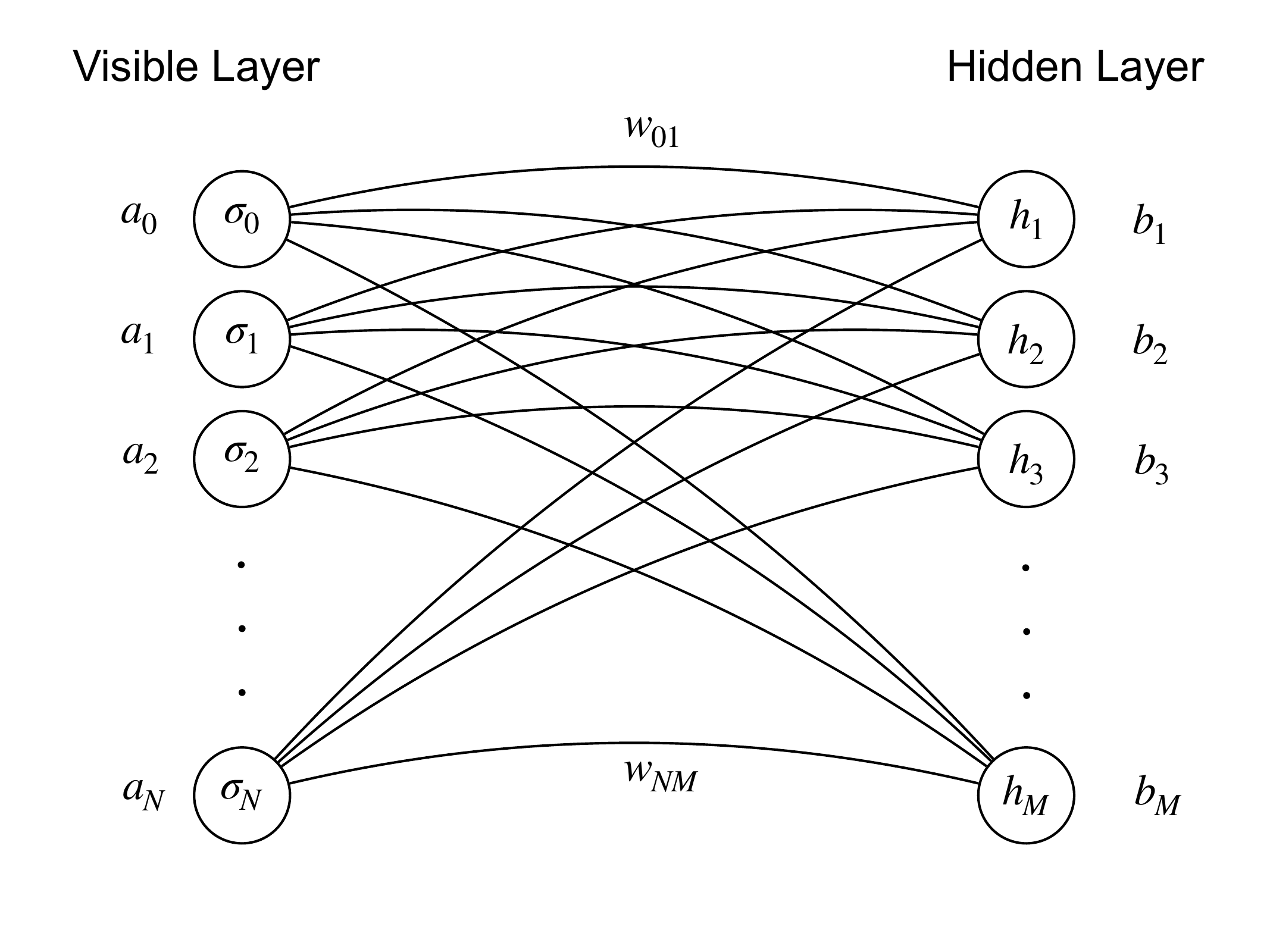}
    \caption{A restricted Boltzmann machine (RBM) architecture applied to the Gaudin magnet. The visible layer, shown on the left, is composed of Ising spin variables $\sigma=\{\sigma_0,\sigma_1,...,\sigma_{N}\}$, describing the eigenstates of Pauli operators $\sigma_j^z$ with eigenvalues $\sigma_j=\pm 1$ at each node $j$. The hidden layer, on the right, is made up of the Ising variables ${h_1,h_2,...,h_M}$. The parameters ${a_0,...,a_{N}}$ and ${b_1,...,b_{M}}$ are the bias weights of the visible and hidden nodes, respectively. The parameters $w_{01},...,w_{NM}$ are weights connecting the visible and hidden nodes.} 
    \label{fig:rbm}
\end{figure}

We represent a quantum state vector with an RBM ansatz \cite{carleo2017solving}. An RBM is a two-layer network model (\textbf{Figure \ref{fig:rbm}}), fully defined by a set of parameters $\mathcal{W}= \{a_j,b_i,w_{ji}\}$, where here we choose the parameters to be complex-valued. The RBM representation of a quantum state $\ket{\psi}$ in the computational basis ($\sigma_j^z$-eigenbasis) $\sigma = \sigma_0,... \sigma_N$ is
\begin{equation}
\ket{\psi}=\sum_{\sigma}C_{\sigma}\ket{\sigma}\propto  \sum_{\sigma}\Psi_{\mathcal W}(\sigma)\ket{\sigma} =: \ket{\mathcal W},
\label{RBMState}
\end{equation}
where the ``$\propto$'' symbol indicates that the RBM representation of a quantum state is generally unnormalized and where $\sigma_j^z\left|\sigma\right>=\sigma_j\left|\sigma\right>$ with $\sigma_j=\pm 1$. The unnormalized amplitude $\Psi_{\mathcal W}(\sigma)$ of a particular spin configuration is given by
\begin{equation}
\Psi_{\mathcal W}(\sigma) = \sum_{\{h_i\}}e^{\sum_ja_j\sigma_j+\sum_ib_ih_i+\sum_{ij}w_{ji}\sigma_jh_i},
\label{RBM}
\end{equation}
where ${a_0,...,a_{N}}$ and ${b_1,...,b_{M}}$ are the bias weights for the visible and hidden nodes, respectively. The visible nodes are associated with Ising variables $\sigma_j=\pm 1$ and the  hidden nodes are assigned values $h_i=\pm 1$. The parameters $w_{ji}$ assign an independent weight to each link between a visible node $j$ and a hidden node $i$. For simplicity, we set the number of hidden nodes to be equal to the number of visible nodes, $\frac{N+1}{M} = 1$. In general, this need not be the case and increasing the number of hidden nodes for a fixed number of visible nodes can increase the expressivity of the RBM. Marginalizing over (tracing out) the hidden-node variables $h_i$ yields a nonlinear function that only takes the variables $\sigma$ as its input and this function is fully determined by the network parameters $\mathcal{W}$.

\subsection{Ground State}
To find the ground state, we minimize the energy expectation value \cite{carleo2017solving}, 
\begin{align}
\begin{split}
&E(\mathcal{W}) = 
\frac{ \bra{\mathcal W}H\ket{\mathcal W}}{\bra{\mathcal W}\ket{\mathcal W}}  \\[0.8ex] 
&=\frac{ \sum_{\sigma,\sigma'}\Psi_{\mathcal W}^*(\sigma)\bra{\sigma}H\ket{\sigma'}\Psi_{\mathcal W}(\sigma')}{\sum_{\sigma''}|\Psi_{\mathcal W}(\sigma'')|^2}  \\[0.8ex] 
&= \sum_{\sigma}\Big(\sum_{\sigma'}\bra{\sigma}H\ket{\sigma'}\frac{ \Psi_{\mathcal W}(\sigma')}{\Psi_{\mathcal W}(\sigma)}\Big)\frac{|\Psi_{\mathcal W}(\sigma)|^2}{\sum_{\sigma''}|\Psi_{\mathcal W}(\sigma'')|^2}  \\
&\approx \Big\langle\sum_{\sigma'}\bra{\tilde{\sigma}}H\ket{\sigma'}\frac{ \Psi_{\mathcal W}(\sigma')}{\Psi_{\mathcal W}(\tilde{\sigma})}\Big\rangle_{{\tilde{\sigma}}} =: \Big\langle E_{\mathrm{local}}\Big\rangle_{{\tilde{\sigma}}}.
\label{MC}
\end{split}
\end{align}
Calculating the exact expectation value would generally involve a sum over $2^{N+1}$ basis states $\sigma$. To avoid an exponential cost in the computation time, we instead evaluate this quantity approximately $\langle ... \rangle_{\tilde{\sigma}}$ via samples $\tilde{\sigma}$ obtained from the Metropolis algorithm \cite{metropolis1953equation}, where the samples are drawn from the probability distribution,
\begin{equation}
\pi(\sigma;\mathcal W)=\frac{|\Psi_{\mathcal W}(\sigma)|^2}{\sum_{\sigma'}|\Psi_{\mathcal W}(\sigma')|^2}.
\label{probDis}
\end{equation}
We use the stochastic reconfiguration method \cite{vicentini2022netket} to minimize the energy subject to variations in the parameters $\mathcal W$. Stochastic reconfiguration exploits information encoded in the quantum geometric tensor (the covariance of the logarithmic derivative of $\Psi_\mathcal{W}$) to precondition the gradient used in stochastic gradient descent (SGD). The energy gradient (force vector) is defined as
\begin{equation}
F_{i} = \frac{\partial E}{\partial \mathcal W_i^*} = \langle E_{\mathrm{local}}\mathcal{O}^{\dagger}_i\rangle_{\tilde{\sigma}} -  \langle E_{\mathrm{local}}\rangle_{\tilde{\sigma}}\langle \mathcal{O}^{\dagger}_{i}\rangle_{\tilde{\sigma}},
\label{Force}
\end{equation}
where the logarithmic derivative of the RBM state vector is defined as
\begin{equation}
\mathcal{O}_{k} = \frac{1}{\Psi_{\mathcal W}(\sigma)} \partial_{\mathcal{W}_k}\Psi_{\mathcal W}(\sigma).
\label{partials}
\end{equation}
The partial derivative in Equation \eqref{Force} is taken with respect to the complex conjugate of the variational parameter, since we are looking for the steepest descent of a real-valued function parameterized by complex-valued variables \cite{hunger2007introduction}. The quantum geometric tensor is defined as
\begin{equation}
S_{ik} = \langle \mathcal{O}^{\dagger}_i\mathcal{O}_{k}\rangle_{\tilde{\sigma}} -  \langle \mathcal{O}^{\dagger}_i\rangle_{\tilde{\sigma}}\langle \mathcal{O}_{k}\rangle_{\tilde{\sigma}}.
\label{Cov}
\end{equation}

\begin{figure}
\centering
    \includegraphics[width=0.95\columnwidth]{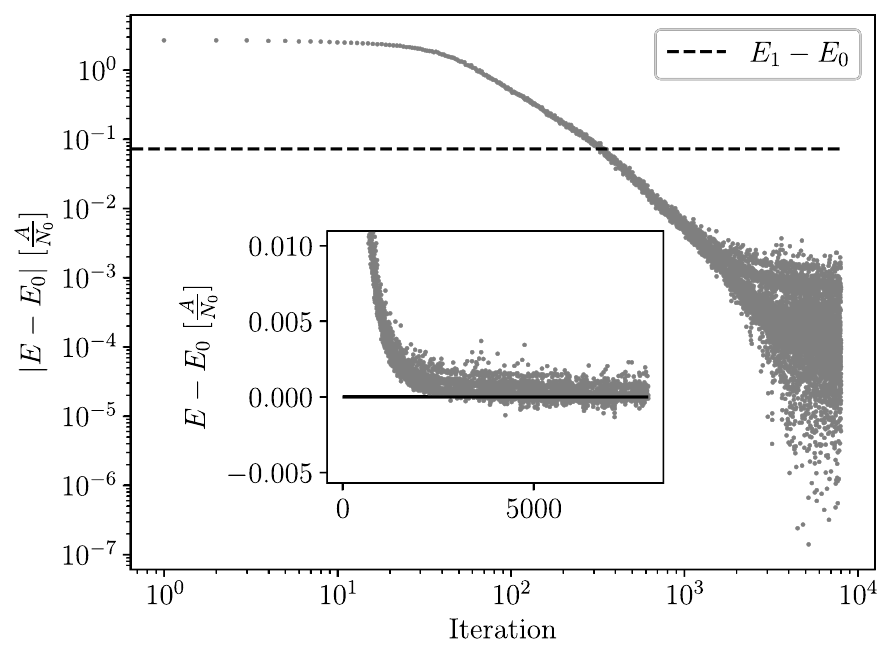}
    \caption{A typical successful run to find the Gaudin-magnet ground state with $N=5$, $B=0.35 A/N_0$, $N_0=3$, and $A_k$ distributed as in Equation \eqref{Ak}. The dashed horizontal line indicates the exact energy $E_1-E_0$ corresponding to the first excited state. Note the logarithmic scale. The subplot displays a zoom-in view near convergence using a linear scale for the energy axis, where the solid horizontal line corresponds to the exact ground state energy.}
    \label{fig:gs}
\end{figure}

At the beginning of an optimization run, a random initial ansatz $\ket{\mathcal W(0)}$ is generated by setting the real and imaginary parts of all variational parameters independently to values obtained from a symmetric Gaussian distribution with standard deviation $\sigma_\mathcal{W}=0.25$. At each iteration $l=1,\ldots,L$, the parameters are updated according to
\begin{equation}
\mathcal{W}(l) = \mathcal{W}(l-1)-\gamma_L S^{-1}_\lambda(l-1)F(l-1),
\label{Update}
\end{equation}
where $\gamma_L$ is the learning rate. Here, $S_\lambda=S+\lambda I$ includes a diagonal shift $\lambda$, a regularization that helps with the matrix inversion \cite{carleo2017solving}. The specific choice of $\lambda$ is given in Sec.~\ref{hyper}, below, along with an explanation of other hyperparameters. See \textbf{Figure \ref{fig:gs}} for a typical successful search for the ground state. The figure shows the estimated energy after each iteration with $L=8000$. This constitutes a single optimization run. At the end of each run, an accurate estimate is obtained for the energy of the final state (for $l=L$) and this is stored. In practice, several runs $p$ are performed ($p=1,2,\ldots,P$) and of those $P$ runs, the run corresponding to the lowest final energy is selected. See \textbf{Algorithm \ref{alg:excited state calculation}}, including the procedure for finding excited states, which we now describe in detail. 

\begin{algorithm}
\caption{Computing the $n^\mathrm{th}$ excited state}
\label{alg:excited state calculation}
\SetKwInOut{Input}{Input}
\SetKwInOut{Output}{Output}
\hrulefill \\
\Input{The $n$ lowest-energy eigenstates in the form of RBMs, \{$\ket{\mathcal {W}^0}, \ket{\mathcal{W}^1}, \ket{\mathcal{W}^2},... \ket{\mathcal{W}^{n-1}}$\},\\ 
and penalty coefficients \{$\beta_0, \beta_1, \beta_2, ... \beta_{n-1}$\}.}
\medskip
\Output{An RBM representation for the $n^\mathrm{th}$ excited state, $\ket{\mathcal{W}^n}$.}
\medskip
\For{optimization runs $p=1,2,3, ... P$}{
Initialize an RBM $\ket{\mathcal W(0)}$\\
    \For{iterations $l=1,2,3,...L$}{
    Estimate the loss function $\tilde{E}_{n}(\mathcal{W})$ \\
    Estimate the gradient $\tilde{F}_{i}(\mathcal{W})$ \\
    Estimate the geometric tensor $S_{ik}(\mathcal{W})$\\
    Regularize: $S_{\lambda,ik}=S_{ik}+\lambda\delta_{ik}$\\
    Update the RBM variational parameters as $\mathcal{W} \leftarrow \mathcal{W}-\gamma S^{-1}_\lambda(\mathcal{W})\tilde{F}(\mathcal{W})$ \\
    }
    Store the final RBM $\ket{\mathcal{W}_\text{final}(p)}$ for each run $p$\\
    Re-estimate $\tilde{E}_{n}(\mathcal{W}_\text{final})$ with more samples and store as $E(p)\leftarrow\tilde{E}_{n}(\mathcal{W}_\text{final})$\\
}
Select the run $p$ with the lowest estimated $E(p)$, giving $\ket{\mathcal{W}^{n}}\leftarrow\ket{\mathcal{W}_\text{final}(p)}$.\\
\hrulefill \\
\medskip
\end{algorithm}

\subsection{Excited States}
Most of the work on neural-network quantum states has focused on the ground state, but a few works have also found approximate excited states by projecting out the previously determined eigenstates \cite{choo2018symmetries, vicentini2022netket}. In general, exactly projecting out the ground state (or any previously determined lower-energy states \cite{jones2019variational, higgott2019variational}) may introduce an exponential cost, associated with the cost of exactly calculating an overlap between the current trial state and the previously determined states. Here, as in the approach of Choo et al.~\cite{choo2018symmetries}, we instead estimate the relevant overlaps with approximate Monte-Carlo sampling. However, rather than directly projecting out the previously calculated eigenstates, we adopt a penalty method \cite{higgott2019variational} to calculate the excited states. In this method, the $n^\mathrm{th}$ excited state is calculated as the ground state of the modified Hamiltonian, 
\begin{equation}
H_n = H + \sum_{j=0}^{n-1}\beta_{j}\frac{ \ket{{\mathcal W^j}}\bra{\mathcal W^j}}{\bra{\mathcal W^j}\ket{\mathcal W^j}}.
\label{varH_ex}
\end{equation}
Here, the terms $\beta_j>0$ are penalty coefficients and $\ket{\mathcal W^j}$ is the approximate $j^\mathrm{th}$ excited state, defined in terms of variational parameters $\mathcal W^j$. The ground state is indexed by $j=0$. Explicitly, we replace the energy expectation value with the new loss function,
\begin{equation}
\tilde{E}_n(\mathcal{W}) =\frac{ \bra{\mathcal W}H\ket{\mathcal W}}{\bra{\mathcal W}\ket{\mathcal W}}+\sum_{j=0}^{n-1}\beta_{j}\frac{ \bra{\mathcal W}\ket{\mathcal W^j}\bra{\mathcal W^j}\ket{\mathcal W}}{\bra{\mathcal W}\ket{\mathcal W}\bra{\mathcal W^j}\ket{\mathcal W^j}}.
\label{varEng_ex}
\end{equation}
After optimization, $\tilde{E}_{n}(\mathcal{W})$ is the approximate energy of the $n^{\mathrm{th}}$ excited state of the original Hamiltonian $H$. The individual penalty coefficients $\beta_{j}$ must be sufficiently large to ensure that the previously calculated eigenstates ($j=0,1,\ldots,n-1$) are all raised in energy (in $H_n$) above $E_n$. However, very large values for $\beta_{j}$ will be detrimental to optimization performance, a well-known phenomenon in penalty-based constrained optimization \cite{bertsekas2014constrained}. Here, our goal is to find a subset of energy eigenstates describing the low-frequency response of the Gaudin magnet to external perturbations. We therefore calculate eigenvalues only up to a maximum cutoff $\omega_\mathrm{max}$ above the ground-state energy. For simplicity, in numerical evaluations we have therefore set all penalty coefficients to a $j$-independent value, 
\begin{equation}
\beta_j = 2 \omega_\textrm{max}. 
\end{equation}
For the specific spin-dynamics results presented below for the Gaudin magnet, we choose $\omega_\mathrm{max}= 0.15\,A/N_0$, which is sufficient to characterize the first five levels with the chosen parameters: $N=5$, $N_0=3$,  $B=0.35A/N_0$, and with the exponential distribution of coupling constants $A_k$ given in Equation \eqref{Ak}.  

\begin{figure}
\centering
    \includegraphics[width=0.95\columnwidth]{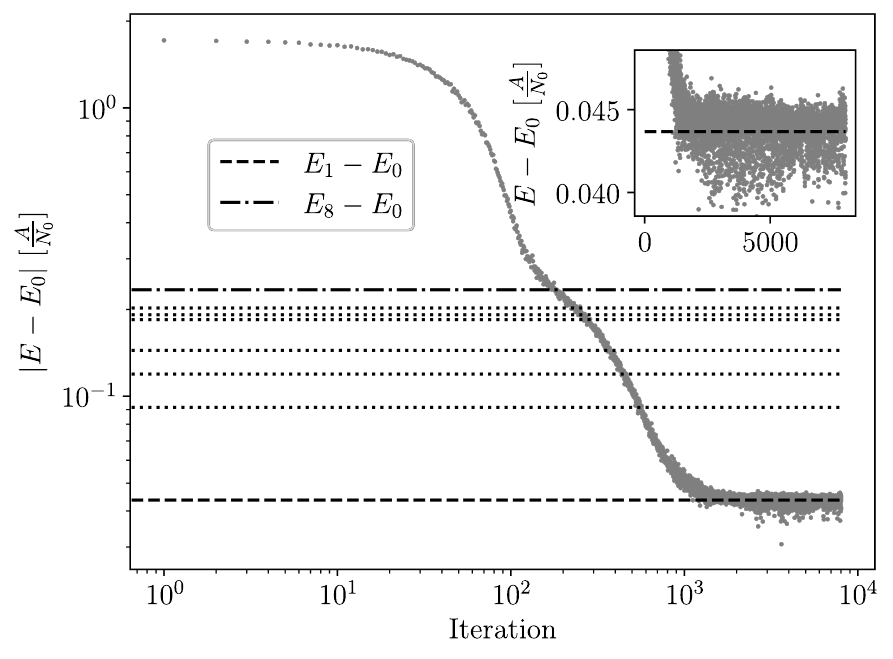}
    \caption{ A typical successful run resulting in the first excited state of the Gaudin magnet with $N=5$. Horizontal lines indicate the exact excitation energies $E_j-E_0$ for the first eight excited states ($j=1,2,\ldots,8$). The inset displays a zoom-in view near convergence with the excitation energy in a linear scale. Convergence slows at an energy close to $E_8$, which strongly suggests the existence of a local minimum at that particular point of the energy landscape. }
    \label{fig:1st}
\end{figure}

\begin{figure}
\centering
\includegraphics[width=0.95\columnwidth]{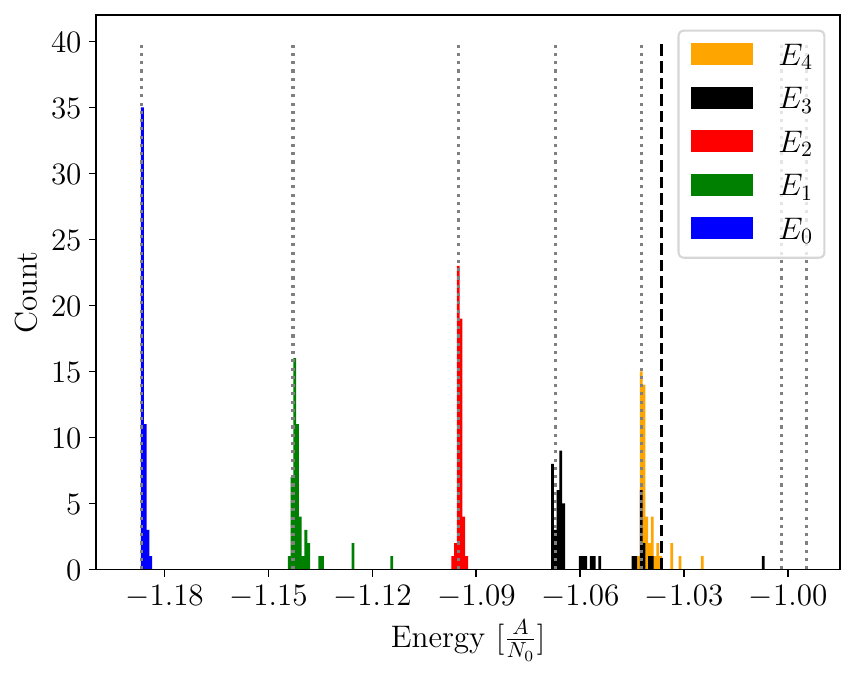}
    \caption{Histogram of approximate energies after 50 runs. The first five energies $E_0, E_1, \ldots, E_4$ lie below a cutoff $E_0+\omega_\mathrm{max}$ (indicated with a vertical dashed line). Other model parameters were as listed in the caption of Figure \ref{fig:gs}. Dotted vertical lines indicate the exact energy levels.}
    \label{fig:histogram}
\end{figure}

The minimization strategy for excited states differs from that of the ground state, since the loss function is modified. While the definition of the quantum geometric tensor $S_{ik}$ remains unchanged, the gradient vector must be modified to (see the Appendix for details):
\begin{align}
\begin{split}
&\tilde{F}_{i}^{n} = \frac{\partial \tilde{E_n}}{\partial \mathcal W_i^*} = \langle E_{\mathrm{local}}\mathcal{O}^{\dagger}_i\rangle_{\tilde{\sigma}} -  \langle E_{\mathrm{local}}\rangle_{\tilde{\sigma}}\langle \mathcal{O}^{\dagger}_{i}\rangle_{\tilde{\sigma}}  \\[0.8ex]
&+  \sum_{j=0}^{n-1}\beta_{j}\left<  \left ( \frac{\Psi_{\mathcal W^j}}{\Psi_{\mathcal W}} - \left<\frac{\Psi_{\mathcal W^j}}{\Psi_{\mathcal W}} \right>_{\tilde{\sigma}} \right ) \mathcal{O}^{\dagger}_{i}\right>_{\tilde{\sigma}} \left<\frac{\Psi_{\mathcal W}}{\Psi_{\mathcal W^j}} \right>_{\tilde{\sigma}_j},
\label{Force_excited}
\end{split}
\end{align}
where the modified energy gradient $\tilde{F}_{i}^{n}$ is evaluated approximately through Metropolis sampling. The samples \{$\tilde{\sigma}$\} are collected based on the distribution $\pi(\sigma; \mathcal W)$ [Equation \eqref{probDis}], whereas \{$\tilde{\sigma}_j$\} are sampled accounting for the variational parameters $\mathcal W^j$, according to $\pi(\sigma; \mathcal W^j)$. See \textbf{Figure \ref{fig:1st}} for a typical successful run resulting in the first excited state. In \textbf{Figure \ref{fig:histogram}} we show a histogram of outcomes for the first five energy levels $E_0,E_1,\ldots,E_4$ after $50$ optimization runs.

\subsection{Hyperparameters and Postselection}\label{hyper}
Here we summarize the chosen model hyperparameters. The learning rate was set to $\gamma_L=0.02$ for all runs. We perform Metropolis sampling with $N_{\tilde{\sigma}}=5000$ samples to estimate the energy loss function [$E(\mathcal{W})$ for the ground state, $\tilde{E}_n(\mathcal{W})$ for the $n^\mathrm{th}$ excited state], the gradient vector $F_i$, and the quantum geometric tensor $S_{ik}$. As described following Equation \eqref{Update}, a diagonal shift $\lambda =  0.01$ is added to the quantum geometric tensor before inversion. The total number of iterations is fixed to $L=8000$.

To further decrease the error of the final approximate ground state, we use the following postselection strategy: 50 independent runs (each consisting of 8000 iterations) were performed with different random seeds, storing the sampled energy and the variational parameters from the last iteration of each run. We then used $10^4$ Metropolis samples (twice as many samples as in the optimization runs) to obtain a more accurate estimate of the energy in each of the 50 runs. We then select the state with the lowest energy (out of the 50 states found from the 50 runs) to be the best approximate ground state. The postselection process for each excited state was the same, after replacing $E(\mathcal{W})$ (the estimated energy alone) with $\tilde{E}_n(\mathcal{W})$ (estimated energy plus penalty terms).

See Algorithm \ref{alg:excited state calculation} for a summary of the $n^\mathrm{th}$ excited state calculation with the number of independent runs $P=50$ and the number of iterations $L=8000$ for both the ground state and the excited states. With the chosen hyperparameters, we found good approximations for the lowest five eigenstates. By comparing to exact diagonalization, we find the error in each of the energy eigenvalues is less than $10^{-3}\,A/N_0$. For all five eigenstates, the state infidelity was bounded by $1-\left|\langle \psi_{\mathrm{exact}} \ket{\psi}\right|^2<2\times 10^{-3}$. 

\subsection{Time Scaling}
\begin{figure}
\centering
    \includegraphics[width=0.95\columnwidth]{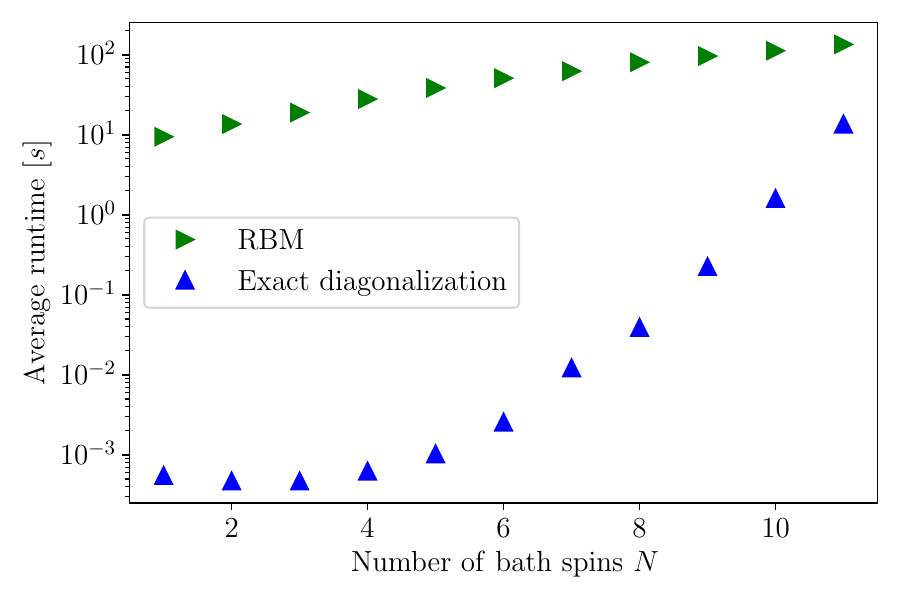}
    \caption{Runtime scaling for the ground state of the Gaudin-magnet Hamiltonian. The runtime for finding the ground state via the RBM-based variational algorithm (blue symbols) is compared to brute-force exact diagonalization (green symbols). For each system size, 50 runs were performed for each of the two algorithms to obtain the average runtime. All numerical runs were performed on the Graham cluster located at the University of Waterloo, Waterloo, ON Canada, with a single CPU and 1024 MB of memory for each run. We also fix both the number of samples and the number of iterations to be 1000. }
    \label{fig:runtime}
\end{figure}

The average runtime (averaged over 50 runs) is shown in \textbf{Figure \ref{fig:runtime}} for a determination of the ground state of the Gaudin magnet for $N=1$ to $N=11$. The runtime for the RBM-based variational algorithm (green triangles) is compared to a brute-force exact diagonalization (blue triangles). Exact diagonalization was performed using the QR algorithm for Hermitian matrices, implemented in NumPy \cite{mushtaq2020numerical}. While the average runtime of brute-force exact diagonalization scales exponentially, the RBM-based variational algorithm with $O(N^2)$ variational parameters (based on the RBM architecture shown in Figure \ref{fig:rbm}) has the potential to scale polynomially. Figure \ref{fig:runtime} suggests that the RBM-based variational algorithm will have a shorter runtime than brute-force exact diagonalization as the system size grows beyond $N\gtrsim 12$. For this comparison, we have fixed the number of samples and the number of iterations. We cannot rule out the possibility that, as the system size grows, the number of samples and iterations required to reach a target error may increase exponentially.  

As we show in Section \ref{sec:Dynamics} below, the ground state of this problem can actually be found through a refined exact diagonalization procedure in a reduced subspace with cost $O(N)$, since the lowest-energy state always lies within a manifold of at most one spin flipped. The comparison here to brute-force exact diagonalization of the entire Hamiltonian is only illustrative and likely not of practical relevance for this particular problem (finding the ground state). However, finding a general excited state (or collection of excited states to describe the system at finite temperature) will typically have exponential cost for direct numerical diagonalization.  

\section{Dynamical response}\label{sec:Dynamics}
\begin{figure*}
\centering
    \includegraphics[width=0.9\textwidth]{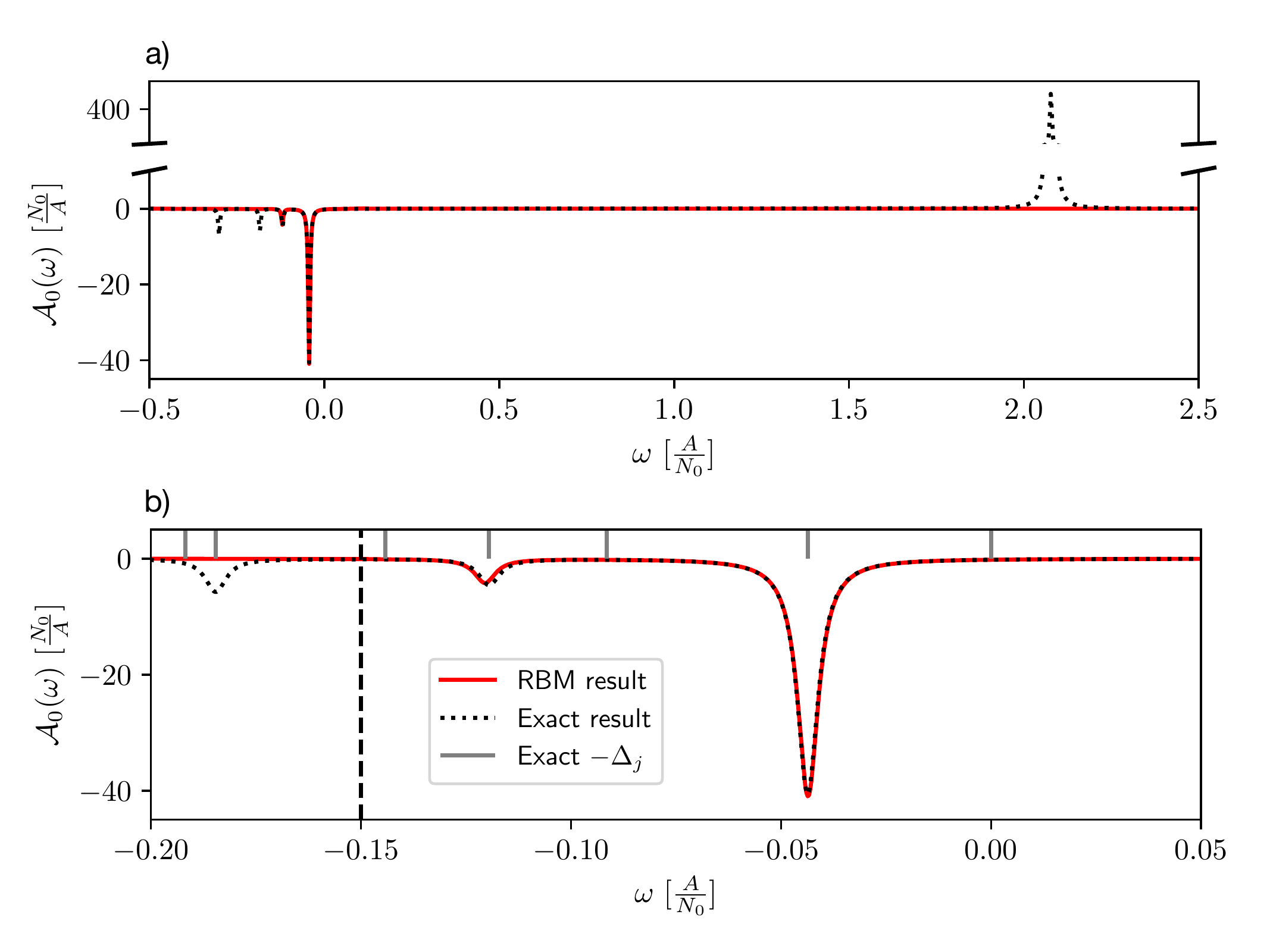}
    \caption{ Central-spin spectral function $\mathcal A_0 (\omega)$ [Equation \eqref{spectral_fun}], with $N=5$ and assuming the exponential distribution of $A_k$ given in Equation \eqref{Ak}. In subplot b), the points $\omega=-\Delta_j$ ($\Delta_j=E_j-E_0$) are indicated with vertical gray lines at the top, and the cutoff frequency $\omega_\mathrm{max} = 0.15 (A/N_0)$ is indicated with a vertical dashed line. We have broadened the peaks with $\gamma = 0.003 A/N_0$.}
    \label{fig:spec_both}
\end{figure*}
\begin{figure*}
\centering
    \includegraphics[width=0.8\textwidth]{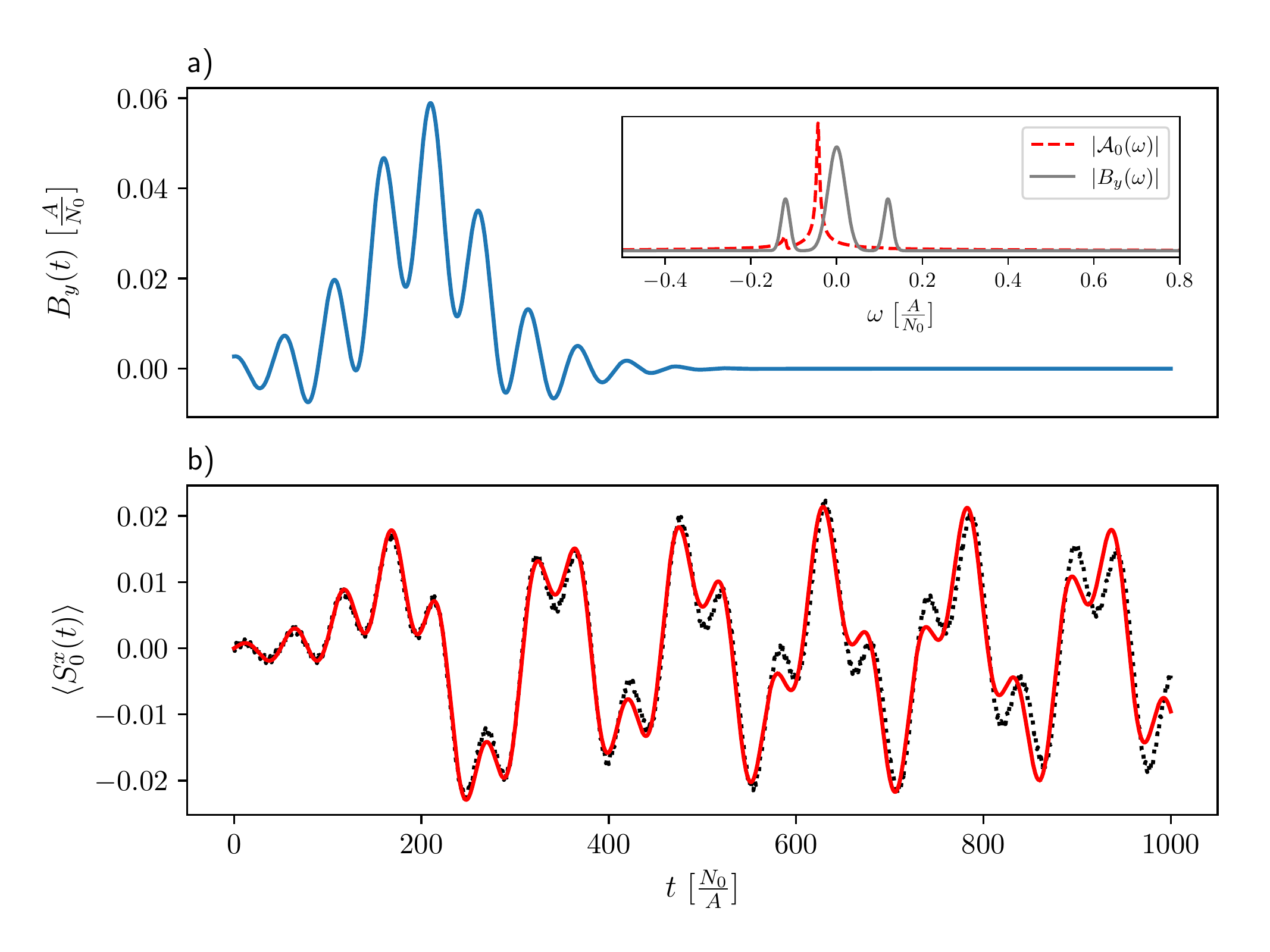}
    \caption{A time-dependent transverse field $B_y(t)$ (subplot a)) drives the central-spin operator $\langle S_0^x(t)\rangle$ (subplot b)) for a Gaudin magnet with $N=5$ environment spins and with an exponential distribution of coupling coefficients $A_k$ given in Equation \eqref{Ak}. In subplot b), the red curve shows the approximate result in linear response calculated from an RBM neural-network ansatz and the dotted black curve is the exact result, obtained from exact numerical integration of the Schr\"odinger equation. The embedded plot in subplot a) shows the spectral weight in the Fourier transform $B_y(\omega)$ relative to the central-spin spectral function $\mathcal{A}_0(\omega)$.}
    \label{fig:linear_xy}
\end{figure*}
Given an accurate representation for both the low-lying spectrum of a many-body quantum system, as well as the low-lying eigenstates, dynamical response functions can be found. In the presence of a time-varying transverse field $\mathbf{B}_\perp(t)=B_x(t)\hat{x}+B_y(t)\hat{y}$ acting on the central spin of the Gaudin magnet, the total Hamiltonian becomes
\begin{equation}
H_\mathrm{tot}(t)=H+V(t),
\end{equation}
with time-dependent perturbation
\begin{eqnarray}
V(t) & = & B_x(t)S_0^x+B_y(t)S_0^y\\
 & = & \frac{1}{2}\left(B_-(t)S_0^++B_+(t)S_0^-\right).
\end{eqnarray}
Here, we have introduced 
\begin{eqnarray}
S_0^\pm & = & S_0^x\pm i S_0^y,\\
B_\pm(t) & = & B_x(t)\pm i B_y(t).
\end{eqnarray}

We assume an initial state $\rho(t_0)$ that is stationary with respect to the Gaudin-magnet Hamiltonian $H$:
\begin{equation}
    \comm{H}{\rho(t_0)}=0.
\end{equation}
This is true for a thermal state or any other statisitical mixture of $H$ eigenstates.  If the initial state is a mixture of non-degenerate Gaudin-magnet eigenstates, the linear response of $\left<S_0^-(t)\right>=\left<S_0^x(t)\right>-i\left<S_0^y(t)\right>$ to $\mathbf{B}_\perp(t)$ is then
\begin{equation}\label{eq:SminusLinearResponse}
\left<S_0^-(t)\right>=-\frac{i}{2}\int_{t_0}^t dt' \left<\comm{\hat{S}_0^-(t-t')}{S_0^+}\right>B_-(t').
\end{equation}
Here, $S_0^-(t)$ evolves under the action of the total Hamiltonian $H_\mathrm{tot}(t)$, while $\hat{S}_0^-(t)=e^{iHt}S_0^- e^{-iHt}$ evolves in the interaction picture under $H$. The average is understood to be taken with respect to the initial state, $\left<\cdots\right>=\mathrm{Tr}\left\{\cdots\rho(t_0)\right\}$. We have also used the following identities, valid for a mixture of non-degenerate $H$ eigenstates: 
\begin{equation}
\left<S_0^-(t_0)\right>=0,
\end{equation}
and 
\begin{equation}
\left<\hat{S}^+_0(t)S^+_0\right>=\left<\hat{S}^-_0(t)S^-_0\right>=0.
\end{equation}
These identities follow directly from the fact that the Gaudin-magnet Hamiltonian commutes with the total $z$-component of spin:
\begin{equation}
\comm{H}{J^z}=0;\quad J^z=\sum_{k=0}^N S_k^z,
\end{equation}
while the operators $S_0^\pm$ couple sectors of different $J^z$. 

Rewriting Equation \eqref{eq:SminusLinearResponse} in terms of Fourier transform variables and taking $t_0\to-\infty$ gives
\begin{equation}
\left<S_0^-(\omega)\right>=\frac{1}{2}\chi^{-+}(\omega)B_-(\omega),
\end{equation}
where the transverse spin susceptiblity is
\begin{equation}
\chi^{-+}(\omega) = -i\int_{-\infty}^\infty dt e^{i\omega t-0^+ |t|}\left<\comm{\hat{S}_0^-(t)}{S_0^+}\right>\theta(t).
\label{chi-+}
\end{equation}
Here, $\theta(t)$ is the Heaviside step function and $0^+$ is a positive infinitesimal. In physical applications, the Gaudin-magnet Hamiltonian will only be an approximate description. Corrections to this description (coupling to a continuum) will lead to a finite decay rate. We account for this effect with the phenomenological replacement
\begin{equation}
0^+\to \gamma.
\end{equation}

We now specialize to the case where the Gaudin magnet is prepared in a non-degenerate ground state, $\rho(t_0)=\left|0\right>\left<0\right|$, where $H\ket{0}=E_0\ket{0}$. Expanding in a complete set of energy eigenstates then gives
\begin{equation}
\chi^{-+}(\omega) = \sum_{j}\left(\frac{\left|\left<j\right|S_0^+\left|0\right>\right|^2}{\omega-\Delta_j+i\gamma}-\frac{\left|\left<j\right|S_0^-\left|0\right>\right|^2}{\omega+\Delta_j+i\gamma}\right),
\label{chi_-+}
\end{equation}
where the $j^\mathrm{th}$ excitation energy is
\begin{equation}
    \Delta_j=E_j-E_0.
\end{equation}
The excitations that can be generated through central-spin spin flips will result in peaks of the central-spin spectral function,
\begin{eqnarray}
\mathcal{A}_0(\omega) & = & -2\mathrm{Im}\chi^{-+}(\omega)\\
 & = & 2\pi\sum_{j,s=\pm}s\left|\left<j\right|S_0^s\left|0\right>\right|^2\delta_\gamma(\omega- s\Delta_j),
\label{spectral_fun}
\end{eqnarray}
where we have introduced a lineshape function
\begin{equation}
    \delta_\gamma(\omega) = \frac{\gamma/\pi}{\omega^2+\gamma^2}.
\end{equation}
For $\gamma\to 0^+$, the lineshape function approaches a Dirac delta function, $\delta_\gamma(\omega)\to\delta(\omega)$. Positive-frequency ($\omega>0$) contribuions to $\mathcal{A}_0(\omega)$ arise from excitations generated through the action of $S_0^+$ on the ground state, while negative-frequency ($\omega<0$) contributions arise from excitations produced through $S_0^-$.

Without loss of generality, we take $B\ge 0$ to define the $+z$ direction of spin. A finite value, $B\ne 0$, favors a ground state with the central spin pointing down, $\ket{\Downarrow}$. When the couplings are negative, $A_k<0$, the environment spins will then favor an orientation along the central spin in the ground state. This fully polarized state is an exact eigenstate of $H$, while for $A_k>0$, the ground state will have mixed character:
\begin{alignat}{3}
    \ket{0} & =  \ket{\Downarrow\downarrow\downarrow\cdots\downarrow}{}\qquad\qquad\qquad& (A_k<0),\\
    \ket{0} & =  \beta_0\ket{\Downarrow\uparrow\uparrow\cdots}+\sum_{k=1}^N\beta_k\ket{\psi_k}\quad & (A_k>0),
\end{alignat}
where
\begin{equation}
    \ket{\psi_k}  \equiv  \ket{\Uparrow\uparrow\uparrow\cdots\downarrow_k\cdots\uparrow}.
\end{equation}

For $A_k<0$, we have
\begin{alignat}{3}
S_0^-\ket{0} & =  0\qquad\qquad &(A_k<0),\\
S_0^+\ket{0} & =  \ket{\Uparrow\downarrow\downarrow\cdots\downarrow}\quad &(A_k<0).
\end{alignat}
This leads to peaks in the spectral function only at positive frequencies $\omega>0$ for $A_k<0$, with amplitudes determined by the overlaps $\left<j\right|\left.\Uparrow\downarrow\downarrow\cdots\downarrow\right>$. In general, we will need detailed knowledge of the many-body eigenstates $\ket{j}$ as well as the excitation energies $\Delta_j$ to accurately determine the spectral function.

For positive coupling, $A_k>0$, we have instead
\begin{alignat}{3}
S_0^-\ket{0} & =  \sum_{k=1}^N\beta_k\ket{\Downarrow\uparrow\uparrow\cdots\downarrow_k\cdots\uparrow}\quad & (A_k>0),\\
S_0^+\ket{0} & =  \beta_0\ket{\Uparrow\uparrow\uparrow\cdots\uparrow}\quad\qquad\qquad & (A_k>0).\label{eq:APositiveOmegaPositive}
\end{alignat}
In contrast with the case of $A_k<0$, the case of $A_k>0$ will lead to peaks at both negative and positive frequencies. Since the fully polarized state in Equation \eqref{eq:APositiveOmegaPositive} is an exact eigensate of $H$ with eigenvalue $(B+\sum_k A_k/2)/2$, there will be a single peak at $\omega\simeq B+\sum_k A_k/2>0$ controlled by the amplitude $\beta_0$, while there will generally be a family of peaks at low negative frequencies $\omega\simeq -\Delta_j$, with amplitudes controlled by $\sum_k\beta_k\left<j\right|\left.\Downarrow\uparrow\uparrow\cdots \downarrow_k\cdots\uparrow\right>$ (see \textbf{Figure \ref{fig:spec_both}} for an example). In what follows, we focus on this case of $A_k>0$.

Based on the discussion here, we see that the zero-temperature central-spin spectral function can, in fact, be calculated numerically exactly with only a polynomial cost. For $A_k<0$, the problem amounts to calculating the excitation energies $\Delta_j$ and overlaps $\left<j\right|\left.\Uparrow\downarrow\downarrow\cdots\downarrow\right>$ after finding the eigenstates $\ket{j}$ through exact diagonalization in the $O(N)$-dimensional subspace of one spin flipped. For the opposite case of $A_k>0$, in addition to calculating the ground state in the subspace of one spin flipped, the associated excitation energies $\Delta_j$ and overlaps $\left<j\right|\left.\Downarrow\uparrow\uparrow\cdots \downarrow_k\cdots\uparrow\right>$ can then be found through exact diagonalization in the $O(N^2)$-dimensional subspace of two spins flipped. In this (zero-temperature) limit, there is likely no practical need for a machine-learning (RBM) approach to dynamics. However, for a finite-temperature thermal state or for initial conditions corresponding to a non-equilibrium configuration described by many spin flips, an efficient alternative to exact diagonalization becomes important. With $n$ flipped spins, the number of states in the subspace (${N \choose n}\sim (N/n)^n$ for $N\gg n\gg 1$) grows exponentially with $n$. While exact-diagonalization necessarily becomes computationally expensive in this limit, the machine-learning (RBM) approach generalizes directly to different initial conditions and has the potential to scale favorably. 

A possible alternative to the RBM method presented here is perturbation theory, but as we show in the following section, perturbation theory fails for this problem in the limit of small-to-moderate $B$ ($B\lesssim A$).  

\subsection{Perturbation theory}

For a sufficiently large $B$, it is possible to find perturbative approximations for the excitation energies and amplitudes, by writing $H=H_0+V_\mathrm{ff}$ with 
\begin{eqnarray}
H_0 & = & \left(B+\sum_k A_k S_k^z\right)S_0^z,\\
V_\mathrm{ff} & = & \frac{1}{2}\sum_kA_k\left(S_0^+S_k^-+S_0^-S_k^+\right).
\end{eqnarray}
An expansion in the flip-flop terms $V_\mathrm{ff}$ then gives:
\begin{eqnarray}
|\beta_0|^2&=&1-\sum_{k=1}^N|\beta_k|^2\\
    & \simeq & 1-\frac{\sum_k A_k^2}{(B+A/2)^2}\\
    & \simeq & 1-\frac{A^2}{2N_0(B+A/2)^2}.
\end{eqnarray}
In the second line above, we have used $\sum_k A_k\simeq A$ for large $N$ and in the third line we have used the specific coupling coefficients given in Equation \eqref{Ak} to evaluate the sum, $\sum_k A_k^2\simeq \int_1^\infty dk A_k^2=A^2/2N_0$. This contribution to the spectral function will dominate ($|\beta_0|^2\simeq 1$) and perturbation theory is valid for calculating this quantity for any non-negative $B$ whenever $N_0\gg 1$. The parameter $|\beta_0|^2$ is the same quantity that gives rise to the long-time saturation in spin coherence describing non-Markovian partial coherence decay for a central-spin system when the interaction is introduced suddenly \cite{coish2004hyperfine}. If only these high-frequency features were of interest, there would typically be no need for a non-perturbative approach (although higher-order corrections in $V_\mathrm{ff}$ will broaden this sharp feature \cite{coish2010free}). The low-frequency contributions to the susceptibility driving the long-time dynamics have a much more stringent condition for perturbation theory to be valid. In particular, due to the possibility for constructive interference of leading-order contributions in perturbation theory, an accurate description of these terms generally requires \cite{coish2004hyperfine}:
\begin{equation}
\left|\sum_k\beta_k\right|\simeq \frac{A/2}{B+A/2}\ll 1.
\end{equation}
This condition is only satisfied for $B\gg A/2$, independent of $N_0$. Outside of this regime, this naïve perturbation theory will fail to accurately describe dynamics. Our focus is now on describing the low-frequency contributions with a non-perturbative method.

We emphasize that the transverse spin susceptibility $\chi^{-+}(\omega)$ gives only the linear response of the central spin to a weak driving field (leading order in $\boldsymbol{B}_\perp$), but an accurate description of the resulting spin dynamics for $B\lesssim A$ requires a nonperturbative calculation to \emph{all orders} in $V_\mathrm{ff}$, even in this limit of a weak driving field.

\subsection{Nonperturbative many-body spin susceptibility}

In this section, we directly apply the RBM ansatz to the problem of finding the zero-temperature dynamical response. While brute-force exact diagonalization is likely to fail for a more general (finite-temperature or nonequilibrium) initial condition due to the exponential growth of the relevant Hilbert space, the machine-learning procedure presented here is directly generalizable to nontrivial initial conditions and has the potential to remain efficient in this regime. In this paper, we simply present a first example involving a small number ($N=5$) of environment spins and we assume a zero-temperature ground state. We leave it to future work to find scaling of the RBM procedure for more general initial conditions. 

To find an accurate approximation for the nonperturbative spin susceptibility, we construct RBM representations of the lowest-energy $j_\mathrm{max}+1$ eigenstates of $H$:   
\begin{eqnarray}
\left|j\right>\simeq \frac{\left|\mathcal{W}^j\right>}{\sqrt{\left<\mathcal{W}^j|\mathcal{W}^j\right>}},\quad j=0,1,\ldots,j_\mathrm{max},
\end{eqnarray}
where $j_\mathrm{max}$ is defined (for a fixed maximum frequency $\omega_\mathrm{max}$) by
\begin{equation}
\Delta_{j_\mathrm{max}}\le \omega_\mathrm{max}<\Delta_{j_\mathrm{max}+1}.
\end{equation}

The relevant matrix elements required to find the susceptibility are then approximated through optimized RBM representations:
\begin{equation}
\left|\left<j\right|S_0^+\left|0\right>\right|^2\simeq \frac{\left<\mathcal{W}^0\right|S_0^-\left|\mathcal{W}^j\right>\left<\mathcal{W}^j\right|S_0^+\left|\mathcal{W}^0\right>}{\left<\mathcal{W}^0|\mathcal{W}^0\right>\left<\mathcal{W}^j|\mathcal{W}^j\right>}, 
\end{equation}
where we rewrite each of the quantum averages in terms of a weighted sum: 
\begin{eqnarray}
\frac{\left<\mathcal{W}^0\right|S_0^-\left|\mathcal{W}^j\right>}{\left<\mathcal{W}^0|\mathcal{W}^0\right>} & = & \left<\sum_{\sigma'}\left<\sigma\right|S_0^-\left|\sigma'\right>\frac{\Psi_{\mathcal{W}^j}(\sigma')}{\Psi_{\mathcal{W}^0}(\sigma)}\right>_0, \label{matrix_ele_1} \\ 
\frac{\left<\mathcal{W}^j\right|S_0^+\left|\mathcal{W}^0\right>}{\left<\mathcal{W}^j|\mathcal{W}^j\right>} & = & \left<\sum_{\sigma'}\left<\sigma\right|S_0^+\left|\sigma'\right>\frac{\Psi_{\mathcal{W}^0}(\sigma')}{\Psi_{\mathcal{W}^j}(\sigma)}\right>_j.
\label{matrix_ele_2}
\end{eqnarray}
Here, we have introduced the notation
\begin{equation}\label{eq:Metropolis}
\left<f(\sigma)\right>_j=\sum_\sigma \pi(\sigma,\mathcal{W}^j)f(\sigma)\simeq \frac{1}{N_{\tilde{\sigma}^j}}\sum_{\tilde{\sigma}^j} f(\tilde{\sigma}^j),
\end{equation}
where $\pi(\sigma,\mathcal{W}^j)$ is given by Equation \eqref{probDis}. On the right-hand side of Equation \eqref{eq:Metropolis}, we have approximated the average by a sample average over $N_{\tilde{\sigma}^j}$ samples $\tilde{\sigma}^j$ that are found in practice through Metropolis sampling.

Once we have the subset of approximate eigenstates in the form of an RBM, we use Metropolis sampling to estimate the matrix elements given in Equation \eqref{matrix_ele_1} and \eqref{matrix_ele_2}. We used $N_{\tilde{\sigma}}=5\times10^6$ samples for the estimate, resulting in a relative statistical error $< 10^{-2}$ for the relevant matrix elements. These approximate matrix elements were then combined with the estimated excitation energies $\Delta_j$ to construct the central-spin spectral function, Equation \eqref{spectral_fun}. See Figure \ref{fig:spec_both} for a comparison of the approximate central-spin spectral function with the result from exact diagonalization. The exact solution accounts for all eigenstates, while the approximate spectral function was calculated using only the lowest five approximate eigenstates. 

\subsection{Linear Response}
\begin{algorithm}
\caption{Computing RBM-based linear response}
\hrulefill \\
\label{alg:linear response}
\SetKwInOut{Input}{Input}
\SetKwInOut{Output}{Output}
\Input{Approximate $n+1$ lowest-energy eigenstates in the form of an RBM ansatz, \{$\ket{\mathcal {W}^0}, \ket{\mathcal{W}^1}, \ket{\mathcal{W}^2}, \newline ... \ket{\mathcal{W}^{n}}$\}, the weak driving field $B_y(t)$, and discrete time points\ \{$t_0, t_1, t_2,... t_\textrm{final}$ \} with fixed time step $\triangle t$.  }
\medskip
\Output{$\langle S^x_0(t) \rangle$ at discrete time points\ \{$t_0, t_1, t_2,... t_\text{final}$ \} }
\medskip
\For{$j=0,1,2,3,...n$}{
Re-estimate the energy $E(\mathcal{W}^{j})$ of the RBM state $\ket{\mathcal{W}^{j}}$ from Equation~\eqref{MC} with $5\times10^6$ samples. \\
Compute $\chi^{xy}(t)$ with the estimated matrix elements given in Equation \eqref{matrix_ele_1} and \eqref{matrix_ele_2}.
}
\For{$t_n = t_0, t_1, t_2, ... t_\textnormal{final}$}{
Compute the discrete convolution integral given in Equation \eqref{linear_res}, as $(\chi^{xy} \ast B_y)(t_n) = \sum_{m=1}^{n} \chi^{xy}(t_m)\cdot B_y(t_{n-m})\cdot \triangle t$.
}
\hrulefill \\
\medskip
\end{algorithm}
 As an illustrative application of the linear-response calculation above, here we consider the real-time dynamics of the central spin in the presence of a slowly-varying transverse field. See \textbf{Algorithm \ref{alg:linear response}} for the steps taken to perform this computation. The spin susceptibility defined in Equation \eqref{chi_-+} directly gives this result. In particular, for a total Hamiltonian $H_\textrm{total} = H + V(t) = H + B_y(t)\cdot S_0^y$, and for a $t=0$ initial ground state of $H$, the linear response is
\begin{equation}
\begin{aligned}
&\left\langle S_0^x(t) \right\rangle = \int_{0}^{t} dt' \chi^{xy}(t-t')B_y(t'),
\label{linear_res}
\end{aligned}
\end{equation}
where
\begin{eqnarray}
\chi^{xy}(t) & = & -i\left<\left[\hat{S}^x(t),S^y\right]\right>\\
    & = & -\frac{1}{2}\mathrm{Re}\left<\left[\hat{S}^-(t),S^+\right]\right>.
\end{eqnarray}
Finally, we choose $B_y(t)$ to have nonvanishing spectral weight at two of the excitation frequencies (inset of \textbf{Figure \ref{fig:linear_xy}}a)). For an illustrative example, we make the specific choice
\begin{equation}
B_y(t) = B_1 g(t-\bar{t},\tau_1)\cos(\Delta_3 t)+B_2 g(t-\bar{t},\tau_2),
\label{ft}
\end{equation}
where $g(x,\sigma_x)$ is a normalized Gaussian envelope:
\begin{equation}
g(x,\sigma_x)=\frac{1}{\sqrt{2\pi}\sigma_x}e^{-\frac{x^2}{2\sigma_x^2}}.
\end{equation}
Here, $B_{1} = B_{2} = 5 \frac{A}{N_0}$, $\bar{t} = 200 \frac{N_0}{A}$, $\tau_{1} = 100 \frac{N_0}{A}$ and $\tau_{2} = 50 \frac{N_0}{A}$. The functional form of $B_y(t)$ is shown in Figure \ref{fig:linear_xy}a) and the resulting linear response $\left<S^x(t)\right>$ is shown in Figure \ref{fig:linear_xy}b). The neural-network (RBM) estimate (solid red curve) accurately reproduces the exact dynamics (black dashed curve) in this linear response regime. The exact result (black dashed line in Figure \ref{fig:linear_xy}b)) was obtained by integrating the exact Schr\"odinger equation directly via the Python package Qutip \cite{johansson2012qutip}, without any linear-response assumption. 

The amplitude of the field $B_y(t)$ is deep in the linear-response regime, so we expect that deviations between the exact and approximate curves shown in Figure \ref{fig:linear_xy}b) are primarily due to errors in the approximate variational (RBM) eigenstates and in the calculations of matrix elements through Metropolis sampling. There may also be some small correction due to the finite-frequency cutoff $\omega_\mathrm{max}=0.15(A/N_0)$ taken in the approximate RBM calculation.

\section{Conclusions}\label{sec:Conclusions}
In this paper, we have applied a neural-network method to calculate approximate low-lying eigenstates and linear response dynamics of an integrable many-body Hamiltonian: the Gaudin magnet. Having an efficient solution for this many-body system could be an important tool in characterizing decoherence sources for qubits that interact with uncontrolled two-level systems. These systems include qubits based on spin, charge, flux, etc., interacting with nuclear spins, coherent charge traps, or paramagnetic impurities. The very same model (the Gaudin magnet) can be used to study the dynamics of nonequilibrium superconductivity since a linear combination of commuting Gaudin-magnet Hamiltonians can be used to represent the Richardson model (BCS Hamiltonian) that drives superconducting pairing dynamics.

This work was motivated in large part by the fact that the Gaudin magnet is a quantum integrable model, admitting a large number of conserved quantities. For such integrable models consisting of $N$ particles, it is well known how to recast the system of $O(e^N)$ linear equations arising from the Schr\"odinger equation in terms of $O(N)$ nonlinear equations, via the algebraic Bethe ansatz. It is not generally known how to efficiently solve these nonlinear equations. This understanding has led us to seek a solution to this dynamics problem via a variational neural-network ansatz (the restricted Boltzmann machine). In contrast with brute-force exact diagonalization in the entire Hilbert space, which is guaranteed to have exponential cost, the neural-network approach taken here relies on a limited (polynomial) number of network parameters and approximate sampling from a subset of states. The neural-network approach therefore has the potential to be efficient. The tradeoff is that this variational neural-network approach has no guarantee of convergence or accuracy, but the first results presented here are promising in this respect.  

We have presented a systematic procedure that could be used to explore the dynamics of the Gaudin magnet or other interesting models. We have further given some illustrative examples based on very small systems (one central spin and $N=5$ environment spins). In this analysis, we have not provided any guarantees of convergence, accuracy, or efficiency (although we have provided direct comparisons with exact results suggesting each of these can be achieved for small systems). In future work it will be important to explore these questions systematically by studying systems of progressively larger size and under conditions that push past the boundaries of what can be done via exact diagonalization. Further improvements can be made by adopting contemporary neural-network quantum state architectures such as autoregressive neural-network models \cite{bennewitz2022neural,carrasquilla2021probabilistic,hibat2020recurrent}. In this architecture, the neural-network quantum state is normalized by construction and sampling can be done more efficiently. 

It remains an interesting open question whether a variational machine-learning approach, as explored here, can easily ``learn'' the symmetries of a many-body problem and exploit these to find an efficient solution.

\appendix
\section{Derivation of the gradient}\label{app:Gradient}
In this Appendix, we provide a detailed derivation of the gradient expression from Equation \eqref{Force_excited}. The variational gradient used to find excited states is derived by differentiating the Monte-Carlo estimated energy including a penalty term (Equation \eqref{varEng_ex}). The derivative is evaluated with respect to the model parameter $\mathcal W$. The first contribution in the gradient comes from the estimated energy alone and has been widely used in existing works \cite{carleo2017solving, bennewitz2022neural, vicentini2022netket}. The derivation here focuses on the second contribution to the gradient from the penalty term.

For simplicity, we assume that only one penalty term is present. The gradient becomes
\begin{align} 
\begin{split}
&\frac{\partial }{\partial \mathcal W^*} \left(\frac{\beta_j |z|^2}{\bra{\mathcal W} \ket{\mathcal W}\bra{\mathcal W^j} \ket{\mathcal W^j}} \right) = \\[0.8ex] 
& \frac{\frac{\partial }{\partial \mathcal W^*} \left (\beta_j |z|^2 \right )}{\bra{\mathcal W} \ket{\mathcal W}\bra{\mathcal W^j} \ket{\mathcal W^j}} +  \frac{\partial }{\partial \mathcal W^*} \left(\frac{1}{\bra{\mathcal W} \ket{\mathcal W}} \right) \frac{\beta_j |z|^2}{\bra{\mathcal W^j} \ket{\mathcal W^j}}. \label{deri_1}
\end{split}
\end{align}
Here, we have introduced the parameter $z = \bra{\mathcal W} \ket{\mathcal W^j}$ and the second term above arises from the normalization factors. The numerator of the first term can be expanded as
\begin{align} 
\begin{split}
&\frac{\partial }{\partial \mathcal W^*} \left(\beta_j |z|^2 \right ) = \beta_j \left ( \frac{\partial z }{\partial \mathcal W^*} z^* + \frac{\partial z^* }{\partial \mathcal W^*} z \right ) =  \beta_j \frac{\partial z^*}{\partial \mathcal W^*} z \\[0.8ex]
& = \beta_j \left ( \sum_{\sigma} \Psi_{\mathcal W^j}(\sigma) \frac{\partial \Psi_{\mathcal W} (\sigma)}{\partial \mathcal W^*} \right) \left (  \sum_{\sigma} \Psi_{\mathcal W^j}^*(\sigma) \Psi_{\mathcal W}(\sigma) \right ), \label{deri_2}
\end{split}
\end{align}
where $\frac{\partial z^* }{\partial \mathcal W^*} = 0$ since the holomorphic function $\Psi_{\mathcal W} (\sigma)$ has the property $\frac{\partial \Psi_{\mathcal W} (\sigma) }{\partial \mathcal W^*} = 0$ \cite{stein2010complex}. Incorporating the normalization terms in the denominator, the first term becomes
\begin{align} 
\begin{split}
\beta_j \left \langle \frac{\Psi_{\mathcal W^j}}{\Psi_{\mathcal W}} \mathcal O^{\dagger}_i \right \rangle_{\tilde{\sigma}} \left \langle  \frac{\Psi_{\mathcal W}}{\Psi_{\mathcal W^j}}  \right \rangle_{\tilde{\sigma}_j}.
\label{deri_3}
\end{split}
\end{align}
The second term in Equation \eqref{deri_1} can be expanded as
\begin{align} 
\begin{split}
&\frac{-\beta_j |z|^2}{\bra{\mathcal W^j} \ket{\mathcal W^j} {\bra{\mathcal W} \ket{\mathcal W}}^2} \left ( \sum_{\sigma} \Psi_{\mathcal W}(\sigma) \frac{\partial \Psi_{\mathcal W} (\sigma)}{\partial \mathcal W^*} \right) \\[0.8ex]
&= - \beta_j \left \langle \frac{\Psi_{\mathcal W^j}}{\Psi_{\mathcal W}} \right \rangle_{\tilde{\sigma}} \left \langle \mathcal O^{\dagger}_i \right \rangle_{\tilde{\sigma}} \left \langle  \frac{\Psi_{\mathcal W}}{\Psi_{\mathcal W^j}}  \right \rangle_{\tilde{\sigma}_j}.
\label{deri_4}
\end{split}
\end{align}
Summing up the two terms in Equation \eqref{deri_3} and \eqref{deri_4}, and then generalizing to multiple penalty terms, we recover the gradient expression in Equation \eqref{varEng_ex}. For the normalized neural-network quantum states in, e.g., Ref.~\cite{bennewitz2022neural}, the gradient expression would only contain the first term in Equation \eqref{deri_3}.

\begin{acknowledgments}
We acknowledge funding from the Natural Sciences and Engineering Research Council (NSERC) and from the Fonds de Recherche--Nature et Technologies (FRQ--NT). 
\end{acknowledgments}


%
\bibliographystyle{apsrev4-2}
\bibliography{Gaudin-ML}

\end{document}